\documentclass[reprint,superscriptaddress,amsmath,amssymb,aps]{revtex4-1}

\usepackage{amsmath}
\usepackage {xcolor}
\usepackage{comment}

\usepackage{mathrsfs}
\usepackage{comment}
\usepackage{amssymb}
\usepackage{graphicx}
\usepackage{subfigure}
\usepackage[colorlinks,
            linkcolor=blue,
            anchorcolor=blue,
            citecolor=blue]{hyperref}
\newtheorem{theorem}{Theorem}
\newtheorem{lemma}{Lemma}

\def\ra{\rangle}
\def\la{\langle}

\usepackage{graphicx}
\usepackage{dcolumn}
\usepackage{bm}

\begin{document}


\title{Coherence-mixedness trade-offs}

\author{Qing-Hua Zhang}
\email[]{qhzhang@csust.edu.cn}

\affiliation{School of Mathematics and Statistics, Changsha University of Science and Technology, Changsha 410114, China}

\author{Shao-Ming Fei}
\email[]{feishm@cnu.edu.cn}
\affiliation{School of Mathematical Sciences, Capital Normal University,
Beijing 100048, China}
\affiliation{Max-Planck-Institute for Mathematics in the Sciences, 04103 Leipzig, Germany}

\begin{abstract}
Quantum coherence constitutes a foundational characteristic of quantum mechanics and is integral to emerging quantum resource theories. However, quantum coherence is severely restricted by environmental noise in general quantum processing, indicated by the loss of information of a quantum system. Such processing can be described by the trade-offs between the coherence and the mixedness. Based on the $l_2$ norm coherence, conditional von Neumann entropy and Wigner-Yanase skew information, we derive basis-independent constraints on the attainable quantum coherence imposed by the mixedness of a quantum state, which generalize the prior basis-dependent relations, provide fundamental insights into the latent coherence resources present within arbitrary quantum systems that undergo decoherence and quantify the inherent limits on extractable coherence imposed by environmental noise.
\end{abstract}

\maketitle

\section{Introduction}
Arisen from the superposition principle in quantum mechanics, the quantum coherence is fundamental to myriad quantum phenomena. Extensive researches have elucidated the indispensable role of quantum coherence across diverse domains including low-temperature physics \cite{horodecki2013fundamental,skrzypczyk2014work,PhysRevX.5.021001,PhysRevLett.113.150402,PhysRevLett.112.030602,doi:10.1073/pnas.1411728112}, nanoscale systems \cite{vazquez2012probing,PhysRevB.84.113415,heinrich2021quantum} and biological processes such as energy transport \cite{Lloyd_2011,frohlich1968long,pupeza2020field}. Recently, a resource-theoretic perspective has emerged by viewing quantum coherence as a physical resource enabling certain informational tasks \cite{PhysRevLett.113.140401,PhysRevLett.113.170401,Korzekwa_2016,PhysRevLett.129.130602,PhysRevLett.125.180603}, which spurs intensive investigations on quantification of coherence \cite{PhysRevA.99.022340,PhysRevA.97.062342,PhysRevLett.119.150405,PhysRevLett.116.150502,PhysRevLett.116.120404,PhysRevA.98.032324,PhysRevLett.115.020403}. Within the resource theory framework, any valid measure of coherence depends on a designated reference basis \cite{PhysRevLett.113.140401}. Famous coherence quantifiers include the $l_p$ norm and relative entropy of coherence \cite{PhysRevLett.113.140401}, as well as skew information \cite{PhysRevLett.113.170401,PhysRevA.95.042337}.

The phenomenon of quantum decoherence, referring to the irreversible loss of information when a quantum system interacts with an external environment, was first elucidated conceptually by Zeh in 1970 \cite{zeh1970interpretation}. To enable quantification of information dynamics, Streltsov $et\ al.$ derived the trade-off relations between the $l_1$ norm quantum coherence measure and the linear entropy mixedness under a designated basis \cite{PhysRevLett.115.020403}. Subsequent investigations have further developed the coherence-mixedness trade-off relations by using diverse coherence quantifiers such as relative entropy of coherence and alternate mixedness measures such as von Neumann entropy \cite{Sun_2023,PhysRevA.92.042101,shao2017quantum,zhang2017estimation,sun2022complementary,CHE2023106794,HU20181,PhysRevA.93.032136}. The study of coherence-mixedness trade-offs provides insight into the degradation of quantum coherence under open system dynamics, with implications for the preservation of coherence as a prerequisite for quantum information processing tasks.

Given the basis-independent nature of measures of mixedness, an intriguing question arises regarding the trade-off relation between coherence and mixedness - can such a relation be characterized in a manner intrinsic to a quantum state itself, without reliance on a particular basis? Fortuitously, quantities such as the average coherence and maximal coherence allow for basis-independent formulations, achievable by optimizing over all bases \cite{PhysRevA.92.042101,LUO20192869}. The ability to express coherence-mixedness trade-offs in a basis-independent way via these optimized coherence quantifiers points to an inherent correlation between coherence and mixedness within the quantum state. This represents a significant conceptual advance, elevating the coherence-mixedness relation beyond a merely relative concept into an intrinsic attribute of a quantum system.

In this paper, we focus on establishing some kinds of trade-off relations for the average (maximal) coherence and mixedness, mainly based on the $l_2$ norm measure of coherence, conditional von Neumann entropy and Wigner-Yanase skew information. The rest of the paper is arranged as follows. In Sec. \ref{sec2}, we first introduce the average coherence and maximal coherence, and present some related properties. In Sec. \ref{sec3} we derive several trade-off relations based on $l_2$ norm coherence, conditional von Neumann entropy and Wigner-Yanase skew information, respectively. Finally, we conclude in Sec. \ref{sec4}.

\section{Average coherence and maximal coherence}\label{sec2}

Let $\rho$ be a quantum state in a Hilbert space with $d$ dimension and $\Pi=\{|i\ra\la i|, i=1,\ \cdots, d\}$ be a von Neumann measurement. As any diagonal density matrix in the basis $\{|i\ra, i=1,\ \cdots, d\}$ is incoherent, the usual and intuitive way to quantify the coherence is to measure the distance between the given state and its closest incoherent states according to different distance norms. The $l_p$ norm-based coherence measure is given by \cite{PhysRevLett.113.140401},
\begin{equation}\label{l1}
C_{l_p}(\rho,\Pi)=\min_{\sigma\in \mathcal{I}}\|\rho-\sigma\|_{l_p}=\left(\sum_{i\neq j}^{d} |\rho_{ij}|^p\right)^{\frac{1}{p}}
\end{equation}
for some suitable positive real number $p$, where $\mathcal{I}$ denotes the set of all incoherent states in the basis $\Pi$. The relative von Neumann entropy of coherence is defined by \cite{PhysRevLett.113.140401},
\begin{equation}\label{relative_entropy_coh}
C_r(\rho,\Pi)=\min_{\sigma\in \mathcal{I}}S(\rho||\sigma)=S(\rho_d)-S(\rho),
\end{equation}
where $S(\rho||\sigma)={\rm tr}(\rho\ln\rho-\rho\ln\sigma)$, $S(\rho)=-{\rm tr}(\rho \ln \rho)$ and $\rho_d=\sum_i \la i|\rho |i\ra |i\ra\la i |$ is the post-measurement state.

In fact, whether the density matrix is diagonal or not under a fixed basis can be directly revealed by the commutation relation between the given density matrix and the non-degenerate observable which equivalently determines a basis set. The coherence measure based on Wigner-Yanase skew information is defined as
\begin{equation}\label{wysi}
C_{sk}(\rho,\Pi)=\sum_{i=1}^{d} I(\rho, |i\ra\la i |)=1-\sum_{i=1}^{d} \la i | \sqrt{\rho} |i\ra^2,
\end{equation}
where $I(\rho, A)=-\frac{1}{2}{\rm tr}[\sqrt{\rho},A]^2$ denotes the Wigner-Yanase skew information with respect to the given observable $A$ \cite{PhysRevA.95.042337,PhysRevA.96.022136,PhysRevA.96.022130,Wigner1963INFORMATION}. It merits emphasis that the Wigner-Yanase skew information has proven to be integral within investigations of quantum critical systems, whereby measures of coherence derived from the Wigner-Yanase information have demonstrated particular efficacy in elucidating both phase transitions and factorization of the ground states in spin chains. The utility of Wigner-Yanase skew information-based coherence elucidates the power of employing information-theoretic quantities for garnering insight into quintessentially quantum phenomena within complex many-body systems at criticality \cite{PhysRevB.90.104431,PhysRevB.93.184428}.

It bears underscoring that quantum coherence exhibits non-invariance under arbitrary unitary operations, yet remains unchanged by incoherent unitary processes, as stipulated by its foundational definition. For the fixed basis set $\Pi=\{|i\ra\la i|, i=1,\ \cdots, d\}$, the maximally coherent pure state is given by $|\psi_d\ra=\frac{1}{\sqrt{d}}\sum_{i=1}^d|i\ra$, for which $C_{l_p}(|\psi_d\ra,\Pi)=\frac{d-1}{d^{p-1}}$, $C_r(|\psi_d\ra,\Pi)=\ln d$ and $C_{sk}(|\psi_d\ra,\Pi)=1-\frac{1}{d}$.

Given the basis-dependence of quantum coherence measures, a natural line of inquiry emerges concerning the extremal coherence values attainable for a fixed quantum state over all reference bases. While the minimal coherence trivially evaluates to zero, the relationship between the average and maximal coherences proves more subtle and revealing. The average coherence provides an insight into the typical coherence exhibited by a state, optimizing over bases to quantify an intrinsic property. Meanwhile, the maximal coherence represents the fundamental limit of extractable coherence from a state under an optimized basis.

An arbitrary reference basis is related to a fixed basis under a unitary transformation. Averaging over all all bases is equivalent to integrate over the unitary orbits of the fixed basis. This is tantamount to integration over the unitary group equipped with the normalized Haar measure, which encapsulates the uniform distribution over unitary operators. The average coherence quantifiers thereby are basis-independent and capture the intrinsic coherence properties of quantum states. For any coherence measure $C$ we have the average coherence and the root mean square average coherence \cite{PhysRevA.92.042101,LUO20192869},
\begin{equation}
C^{\mathcal{U}}(\rho)=\int_{\mathcal{U}} C(\rho,U\Pi U^\dagger) dU,
\end{equation}
\begin{equation}
{\rm rms}C^{\mathcal{U}}(\rho)=\left [ \int_{\mathcal{U}} C^2(\rho,U\Pi U^\dagger) dU\right ] ^{\frac{1}{2}},
\end{equation}
where $dU$ denotes the normalized Haar measure on the unitary group $\mathcal{U}$, and $U\Pi U^\dagger=\{U|i\ra\la i |U^\dagger,\, i=1,\ 2,\cdots,\ d \}$. Note that the convexity of the function $f(x)=x^2$ implies the relation $C^{\mathcal{U}}(\rho)\leqslant {\rm rms}C^{\mathcal{U}}(\rho)$.

Note that it is very hard to formula the average coherence based on the $l_p$ norm coherence measure for general $p$. Fortunately, the root mean square average coherence of the $l_2$ norm coherence can be evaluated analytically. We rewrite the square of $l_2$ norm coherence as
\begin{equation}
C_{l_2}^2(\rho,\Pi)={\rm tr}\rho^2-\sum_{i=1}^{d} \la i | \rho |i\ra^2.
\end{equation}
Denote the flip operator by $F=\sum_{ij}|ij\ra\la ji|$. We obtain
$$
\begin{aligned}
&\sum_{i=1}^d \int_{\mathcal{U}}\left\langle i\left|U^{\dagger} {\rho} U\right| i\right\rangle^2 d U \\
=&\sum_{i=1}^d \int_{\mathcal{U}} \operatorname{tr}\left(\left(U|i\rangle\langle i| U^{\dagger}\right)^{\otimes 2} {\rho}^{\otimes 2}\right) d U \\
=&\sum_{i=1}^d \operatorname{tr}\left(\int_{\mathcal{U}}\left(U|i\rangle\langle i| U^{\dagger}\right)^{\otimes 2} d U\right) {\rho}^{\otimes 2} \\
=&\sum_{i=1}^d \operatorname{tr} \frac{{I}+{F}}{d(d+1)} {\rho}^{\otimes 2} \\
=&\frac{1+\operatorname{tr} {\rho}^2}{d+1},
\end{aligned}
$$
where $I$ is the identity, and in the above derivation we have used the relation \cite{PhysRevA.40.4277}
$$
\int_{\mathcal{U}}\left(U|j\rangle\langle j| U^{\dagger}\right)^{\otimes 2} d U=\frac{I+F}{d(d+1)}
$$
and the fact $\operatorname{tr} (F\rho\otimes \rho)=\operatorname{tr}(\rho^2)$. Thus, we have the following lemma.

\begin{lemma}
 The root mean square average coherence based on the $l_2$ norm coherence is given by \cite{PhysRevA.92.042101}
\begin{equation}\label{rmsuc}
{\rm rms}C_{l_2}^{\mathcal{U}}(\rho)=\left(\frac{d{\rm tr}\rho^2-1}{d+1}\right)^\frac{1}{2}.
\end{equation}
\end{lemma}

The above quantity ${\rm rms}C_{l_2}^{\mathcal{U}}(\rho)$ has the following basic properties:
\begin{itemize}
\item[(a)] $0\leqslant {\rm rms}C_{l_2}^{\mathcal{U}}(\rho)\leqslant (\frac{d-1}{d+1})^\frac{1}{2}$. The minimal value is archived if and only if $\rho=\frac{I}{d}$ and the maximal value is archived if and only if $\rho$ is a pure state.
\item[(b)] ${\rm rms}C_{l_2}^{\mathcal{U}}(\rho)^2$ is convex of $\rho$.
\item[(c)] ${\rm rms}C_{l_2}(\rho)$ is unitarily invariant, that is, ${\rm rms}C_{l_2}(U\rho U^\dagger )={\rm rms}C_{l_2}(\rho)$ for any unitary matrix $U$.
\end{itemize}

The average coherence of the relative von Neumann entropy is of the following form \cite{PhysRevA.92.042101,LUO20192869,PhysRevA.49.668},
\begin{equation}\label{acr}
C_r^{\mathcal{U}}(\rho)=L_d-\left[S(\rho)-Q(\rho)\right],
\end{equation}
where $L_d=\frac{1}{2}+\frac{1}{3}+\dots+\frac{1}{d}$ and $Q(\rho)=-\sum_{k=1}^d\left(\prod_{k\neq l}\frac{\lambda_k}{\lambda_k-\lambda_l}\right)\lambda_k\ln \lambda_k$ with $\lambda_k$ the $k$th eigenvalue of $\rho$.

The average coherence of the Wigner-Yanase skew information is of the form \cite{LUO20192869},
\begin{equation}\label{acsk}
C_{sk}^{\mathcal{U}}(\rho)=\frac{d-(\operatorname{tr}\sqrt{\rho})^2}{d+1}.
\end{equation}
Indeed, $C_{sk}^{\mathcal{U}}(\rho)$ has similar properties to ${\rm rms}C^{\mathcal{U}}(\rho)$: $0\leqslant C_{sk}^{\mathcal{U}}(\rho)\leqslant \frac{d-1}{d+1}$; $C_{sk}^{\mathcal{U}}(\rho)$ is convex of $\rho$; and $C_{sk}^{\mathcal{U}}(\rho)$ is unitarily invariant. As an average coherence $C_{sk}^{\mathcal{U}}(\rho)$ can be interpreted as an exact uncertainty relation concerning all the von Neumann measurements, as well as an intrinsic measure of quantum uncertainty of $\rho$.

The maximum coherence of a state is obtained by maximizing over all $\Pi$s \cite{LUO20192869},
\begin{equation}
C^{\operatorname{max}}(\rho)=\max_{\Pi} C(\rho,\Pi).
\end{equation}
The maximum coherence of the $l_2$ norm, the relative von Neumann entropy and the Wigner-Yanase skew information are given by, respectively \cite{yu2016total,LUO20192869,CHE2023106794},
\begin{equation}\label{l2mc}
C_{l_2}^{\operatorname{max}}(\rho)=(\operatorname{tr}\rho^2-\frac{1}{d})^\frac{1}{2},
\end{equation}
\begin{equation}\label{rmc}
C_{r}^{\operatorname{max}}(\rho)=\ln d-S(\rho),
\end{equation}
\begin{equation}\label{wysimc}
C_{sk}^{\operatorname{max}}(\rho)=1-\frac{1}{d}\left(\operatorname{tr} \sqrt{\rho}\right)^2.
\end{equation}

We need to point out that all the maximal coherences above are convex of $\rho$ and are zero if and only if $\rho=I/{d}$. It is interesting to note that the ratio of the average coherence and the maximal coherence tends to 1 when $d\rightarrow \infty$,
\begin{equation*}
\frac{{\rm rms}C_{l_2}^{\mathcal{U}}(\rho)}{C_{l_2}^{\operatorname{max}}(\rho)}=\left(\frac{d+1}{d}\right)^{1/2}\rightarrow 1,
\end{equation*}
\begin{equation*}
\frac{C_{r}^{\mathcal{U}}(\rho)}{C_{r}^{\operatorname{max}}(\rho)}=\frac{Q(\rho)+L_d-S(\rho)}{\ln d-S(\rho)}\rightarrow 1,
\end{equation*}
\begin{equation*}
\frac{C_{sk}^{\mathcal{U}}(\rho)}{C_{sk}^{\operatorname{max}}(\rho)}=\frac{d}{d+1}\rightarrow 1.
\end{equation*}
Roughly speaking, for high-dimensional quantum systems the coherence of a state approaches to the maximal values for nearly all the choices of reference basis. This parallels the phenomenon in bipartite systems whereby almost all pure states exhibit maximal entanglement in the high-dimensional limit.

\section{Complementarity between coherence and mixedness}\label{sec3}

The mixedness of a quantum state characterizes the disorder of the quantum system. The quantification of quantum mixedness can be formulated through entropic functions of quantum states, including the linear and von Neumann entropies, which characterize the informational content of quantum systems. Additionally, the quantum fidelity, which measures the overlap between quantum states, provides a useful metric for assessing the coherence present within a given quantum system. The normalized linear entropy mixedness $M_l(\rho)$, von Neumann entropy mixedness $S(\rho)$ and geometric mixedness $M_g(\rho)$ for arbitrary d-dimensional quantum state $\rho$ are defined by, respectively \cite{PhysRevA.70.052309},
\begin{equation}\label{nlem}
 M_l(\rho)=\frac{d}{d-1}\left(1-\operatorname{tr} \rho^2\right),
 \end{equation}
 \begin{equation}\label{em}
 S(\rho)=-\operatorname{tr}(\rho \ln \rho),
  \end{equation}
   \begin{equation}\label{gm}
M_g(\rho)=F(\rho, I / d)=\frac{1}{d}(\operatorname{tr} \sqrt{\rho})^2
\end{equation}
where $F(\rho,\sigma)=(\operatorname{tr}\sqrt{\sqrt{\sigma}\rho\sqrt{\sigma}})^2$ is the fidelity.

It is important to note that while coherence measures depend on the choice of basis, the maximal and average coherences are basis-independent. Basis-independent complementary relations between coherence and mixedness would provide fundamental insights into the intrinsic resources present in a quantum state.

\begin{theorem}\label{l2th}
For any arbitrary quantum system $\rho$ in $d$ dimensional Hilbert space $H_d$, the $l_2$ norm-based root mean square average coherence ${\rm rms}C_{l_2}^{\mathcal{U}}(\rho)$ and $l_2$ norm-based maximal coherence $C_{l_2}^{\operatorname{max}}(\rho)$ are respectively restricted by the normalized linear entropy mixedness $M_l(\rho)$ through the following equalities,
\begin{equation}
\frac{d+1}{d-1}\left[{\rm rms}C_{l_2}^{\mathcal{U}}(\rho)\right]^2+M_l(\rho)=1,
\end{equation}
\begin{equation}
\frac{d}{d-1}\left[C_{l_2}^{\operatorname{max}}(\rho)\right]^2+M_l(\rho)=1.
\end{equation}
\end{theorem}

Theorem \ref{l2th} can be directly obtained by combining (\ref{nlem}), (\ref{rmsuc}) and (\ref{l2mc}). From Theorem \ref{l2th} the trade-off inequality holds naturally for any given basis $\Pi$: $\frac{d}{d-1}[C_{l_2}(\rho,\Pi)]^2+M_l(\rho)\leqslant 1$, which is in consistent with the Theorem 1 in Ref. \cite{CHE2023106794}. Here, our result provides another proof of this trade-off relation.

Following the fact that \cite{PhysRevA.92.042101},
$C_{l_1}(\rho,\Pi)\leqslant \sqrt{d(d-1)}C_{l_2}(\rho,\Pi)$,
we obtain $$
C_{l_1}^{\operatorname{max}}(\rho)\leqslant \sqrt{d(d-1)}C_{l_2}^{\operatorname{max}}(\rho).
$$
As the maximal coherences of $l_1$ and $l_2$ norm may be obtained from different bases,
our theorem provides a simple proof for the trade-off relation given in \cite{PhysRevA.91.052115},
$$\left[\frac{C_{l_1}^{\operatorname{max}}(\rho)}{d-1}\right]^2+M_l(\rho)\leqslant 1.$$
In addition, according to the fact $C^{\mathcal{U}}(\rho)\leqslant {\rm rms}C^{\mathcal{U}}(\rho)$,  the following trade-off holds naturally between $l_2$ norm-based average coherence $C_{l_2}^{\mathcal{U}}(\rho)$ and the normalized linear entropy mixedness $M_l(\rho)$:
$$
\frac{d+1}{d-1}C_{l_2}^{\mathcal{U}}(\rho)^2+M_l(\rho)\leqslant 1.
$$

The quantitative investigation of wave and particle properties in multi-path interferometry was pioneered by D\"urr in \cite{PhysRevA.64.042113}. The $d$ different paths are denoted by the computational basis $\{|i\ra, i=1,2,\cdots,d\}$, corresponding to von Neumann measurement $\Pi=\{|i\ra\la i|, i=1,2,\cdots,d\}$. Passing through the beam splitter, the output state can be expressed in the computational basis as,
$\rho=(\sum_{i=1}^d|i\ra\la i|)\rho(\sum_{j=1}^d|j\ra\la j|)=\sum_{ij}^d\la i|\rho|j\ra |i\ra\la j|$ \cite{PhysRevA.64.042113}. The wave-particle duality established by D\"urr is given by
$$
P_{\mathrm{D}}^2(\rho, \Pi)+V_{\mathrm{D}}^2(\rho, \Pi) \leqslant 1,
$$
where the predictability and visibility are
$$
\begin{gathered}
P_{\mathrm{D}}^2(\rho, \Pi)=\frac{d}{d-1} \sum_{i=1}^d\left(\langle i|\rho| i\rangle-\frac{1}{d}\right)^2, \\
V_{\mathrm{D}}^2(\rho, \Pi)=\frac{d}{d-1} \sum_{i \neq j}|\langle i|\rho| j\rangle|^2,
\end{gathered}
$$
respectively. Furthermore, D\"urr established the following complementarity relation,
$$
\operatorname{tr} \rho^2=\frac{1}{d}+\frac{d-1}{d}\left[P_{\mathrm{D}}^2(\rho, \Pi)+V_{\mathrm{D}}^2(\rho, \Pi)\right],
$$
which can be regarded as a triality relation called wave-particle-mixedness triality \cite{Fu_2022}.
After simple manipulation, this triality can be transformed into a form which is mathematically equivalent to our triality relation, namely,
$$
\left[C_{l_2}^{\operatorname{max}}(\rho)\right]^2=\frac{d-1}{d}\left[P_{\mathrm{D}}^2(\rho, \Pi)+V_{\mathrm{D}}^2(\rho, \Pi)\right].
$$
Our Theorem \ref{l2th} implies the wave-particle-mixedness triality \cite{Fu_2022},
\begin{equation*}
P_{\mathrm{D}}^2(\rho, \Pi)+V_{\mathrm{D}}^2(\rho, \Pi)+M_l(\rho)=1.
\end{equation*}

\begin{theorem}\label{retorpy}
For any $d$ dimensional quantum state $\rho$, the trade-off relation between the relative von Neumann entropy-based maximum (average) coherence $C_{r}^{\operatorname{max}}(\rho)$ $(C_{r}^{\mathcal{U}}(\rho))$ and von Neumann entropy mixedness $S(\rho)$ is given by
\begin{equation}
{C_{r}^{\mathcal{U}}(\rho)+[S(\rho)-Q(\rho)]}={L_d},
\end{equation}
\begin{equation}
{C_{r}^{\operatorname{max}}(\rho)+S(\rho)}={\ln d}.
\end{equation}
\end{theorem}

Theorem 2 can be proved by using the equalities (\ref{em}), (\ref{acr}) and (\ref{rmc}). The quantum subentropy $Q(\rho)$ is a tight lower bound on the accessible information of the pure-state ensembles and is never greater than the von Neumann entropy $S(\rho)$ \cite{PhysRevA.49.668,10.1063/1.4882935}. Note that the quantity $S(\rho)-Q(\rho)$ denotes a kind of mixedness, for which the maximum $L_d$ is achieved for the maximally mixed state and the minimum $0$ is achieved for pure states. One can directly derive that the average coherence $C_{r}^{\mathcal{U}}$ obeys ${C_{r}^{\mathcal{U}}(\rho)+S(\rho)}\leqslant {\ln d}$.
From Theorem \ref{retorpy}, we also have that the trade-off inequality between the measure of coherence and the mixedness holds naturally for any given basis $\Pi$:
$C_{r}(\rho,\Pi)+S(\rho)\leqslant \ln d.$

\begin{theorem}\label{wyskcohe}
For any $d$ dimensional quantum state $\rho$, the Wigner-Yanase skew information-based average coherence and the maximum coherence $C_{sk}^{\operatorname{max}}(\rho)$ are restricted by the geometric mixedness $M_g(\rho)$,
\begin{equation}
\frac{d+1}{d}C_{sk}^{\mathcal{U}}(\rho)+{M_g(\rho)}=1.
\end{equation}
\begin{equation}
C_{sk}^{\operatorname{max}}(\rho)+M_g(\rho)=1.
\end{equation}
\end{theorem}

Theorem \ref{wyskcohe} is derived from the equalities (\ref{gm}), (\ref{acsk}) and (\ref{wysimc}). It implies that the trade-off inequality between the measure of coherence $C_{sk}(\rho,\Pi)$ and the geometric mixedness $M_g(\rho)$ holds naturally for any given basis $\Pi$ \cite{CHE2023106794}, $C_{sk}(\rho,\Pi)+M_g(\rho)\leqslant 1$.

The geometric coherence is of the form $C_g(\rho)=1-\min_{\sigma\in \mathcal{I}} F(\rho,\sigma)$, where $F(\rho,\sigma)$ is the fidelity of $\rho$ and $\sigma$ \cite{PhysRevLett.115.020403}. According to the fact $C_g(\rho)\leqslant C_{sk}(\rho)$, we obtain $C_g(\rho)+M_g(\rho)\leqslant 1$.
Thus our Theorem \ref{wyskcohe} covers the complementarity of the geometric coherence and the geometric mixedness presented in \cite{CHE2023106794}.

\begin{figure}[tbp]
	 \subfigure[$d=2$]
	 {
	 \centering
	\label{figex1a}
		\includegraphics[width=4.1cm]{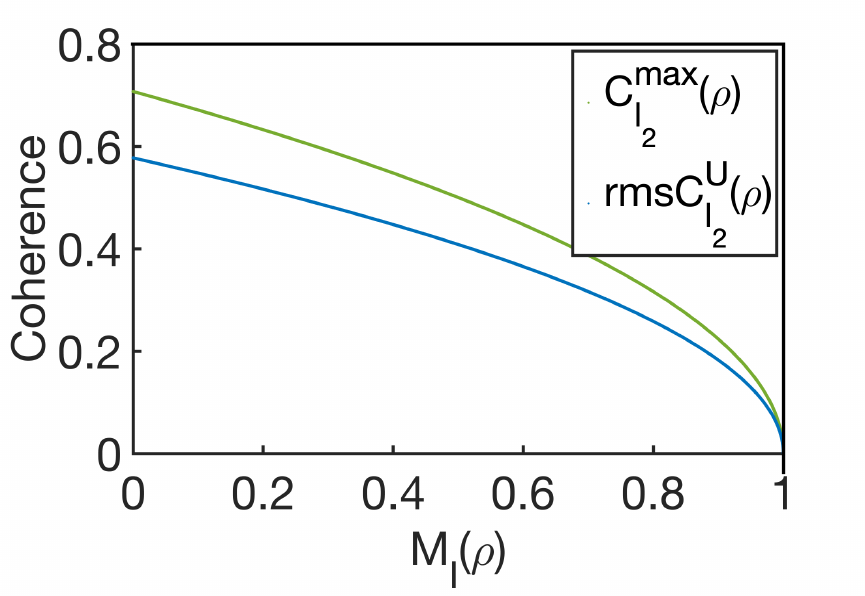}}
		\subfigure[$d=3$]
		{
		\centering
		\label{figex1b}
		\includegraphics[width=4.1cm]{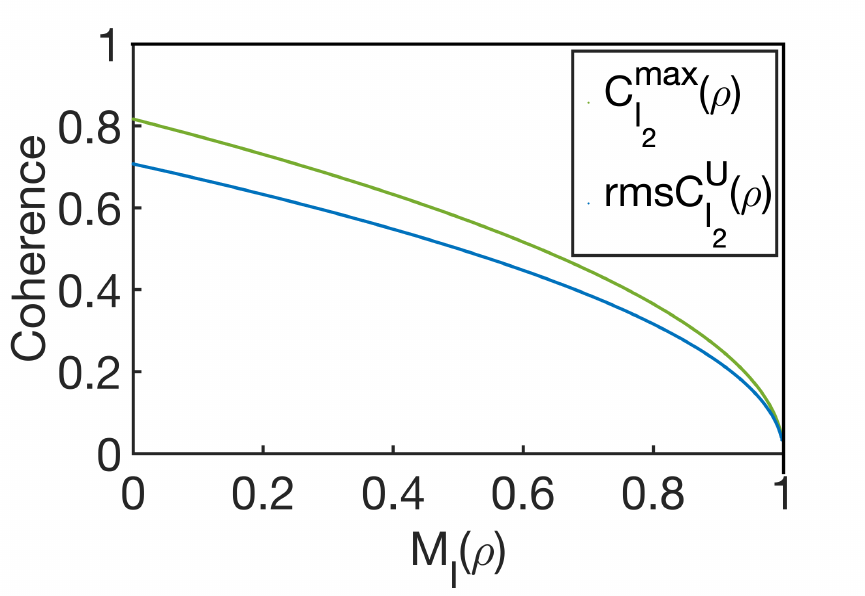}}
		\subfigure[$d=4$]
		{
		\centering
		\label{figex1b}
		\includegraphics[width=4.1cm]{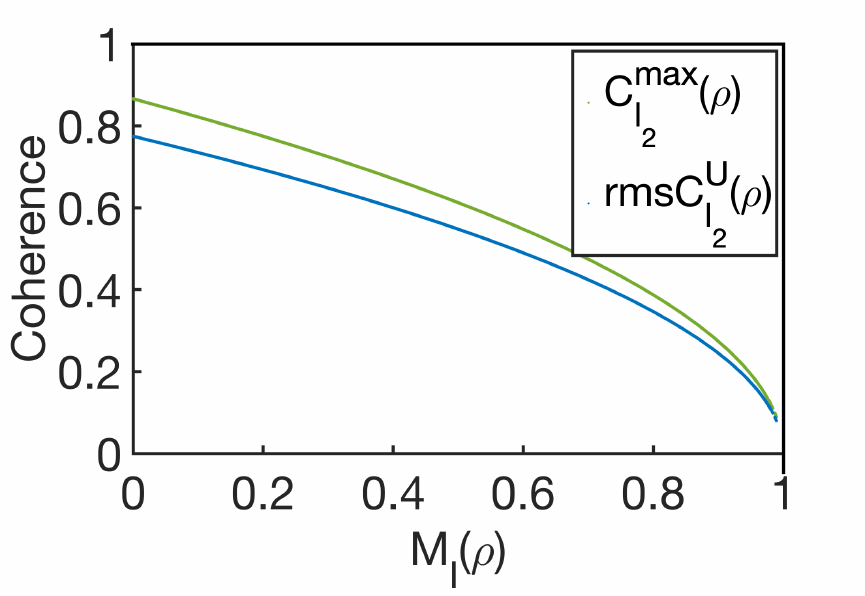}}
		\subfigure[$d=5$]
		{
		\centering
		\label{figex1b}
		\includegraphics[width=4.1cm]{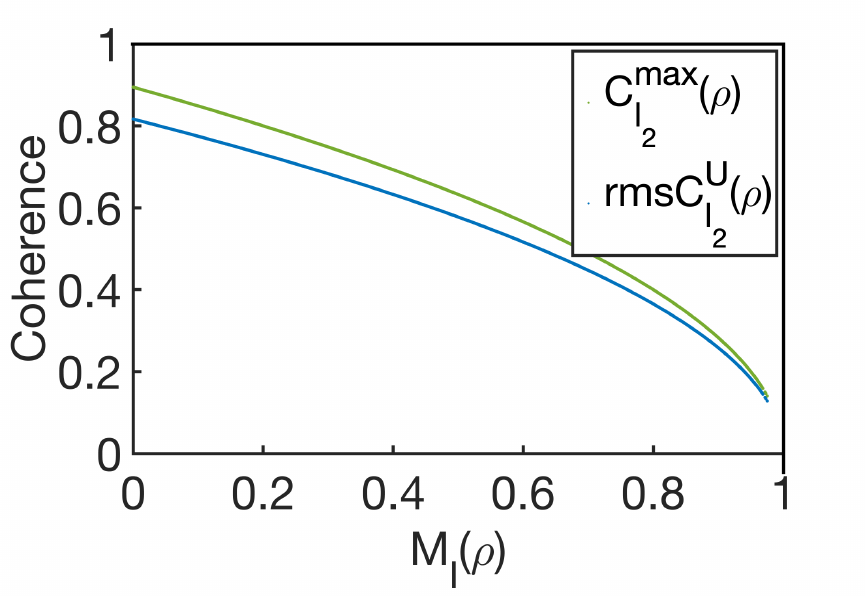}}
 \caption{$10^5$ random states are generated to illustrate the trade-offs between the $l_2$ norm-based maximal (average) coherence and the linear entropy mixedness.}
 \label{fig1}
 \end{figure}

\begin{figure}[tbp]
	\subfigure[$d=2$]{
	 \centering
	\label{figex1a}
		\includegraphics[width=4.1cm]{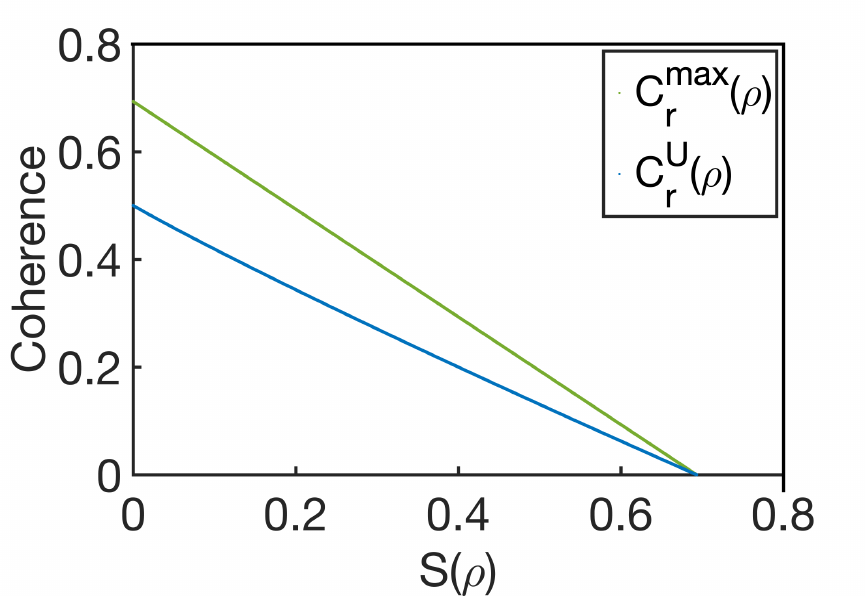}}	
		\subfigure[$d=3$]{
		\centering
		\label{figex1b}
		\includegraphics[width=4.1cm]{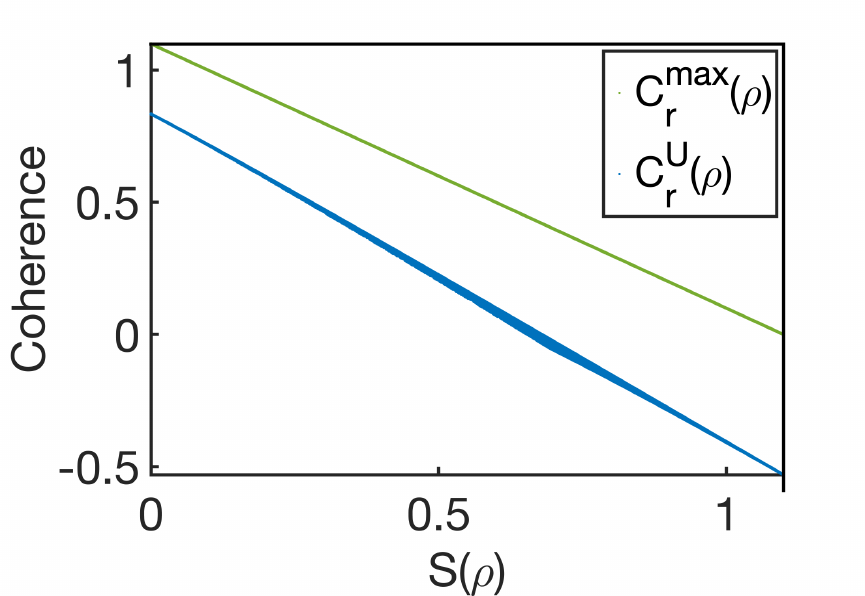}}	
		\subfigure[$d=4$]{
		\centering
		\label{figex1b}
		\includegraphics[width=4.1cm]{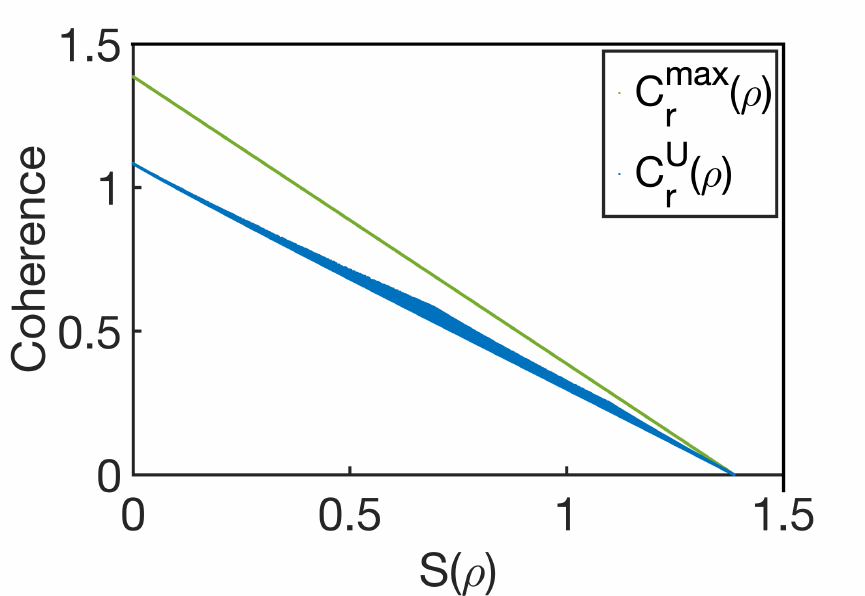}}	
		\subfigure[$d=5$]{
		\centering
		\label{figex1b}
		\includegraphics[width=4.1cm]{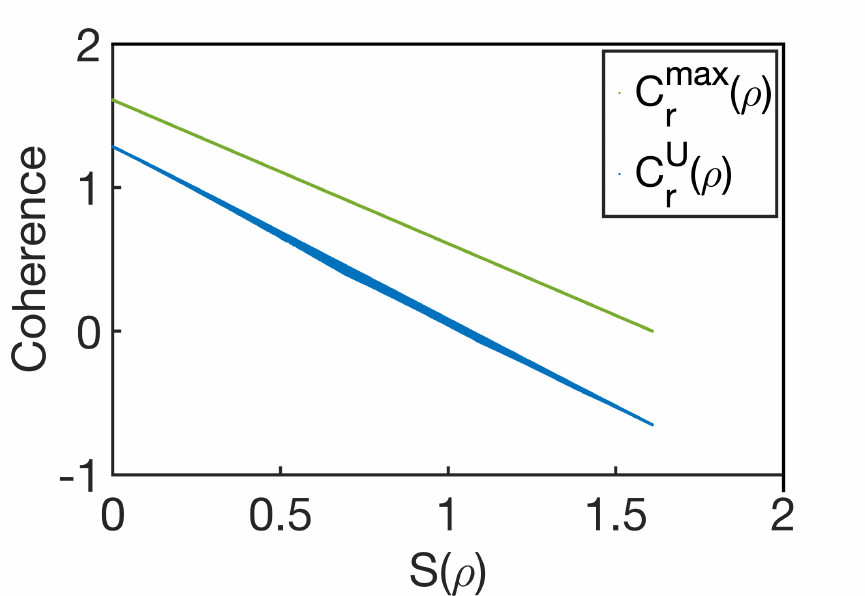}}
 \caption{$10^5$ random states are generated to illustrate the trade-offs between the von Neumann entropy-based maximal (average) coherence and the von Neumann entropy mixedness.}
 \label{fig2}
 \end{figure}

\begin{figure}[tbp]
	\subfigure[$d=2$]{
	 \centering
	\label{figex1a}
		\includegraphics[width=4.1cm]{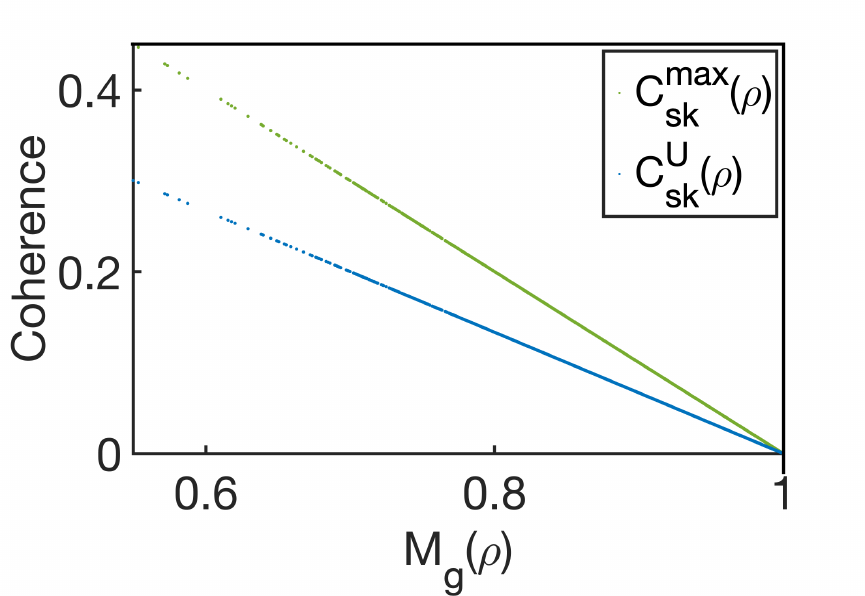}}
		\subfigure[$d=3$]{
		\centering
		\label{figex1b}
		\includegraphics[width=4.1cm]{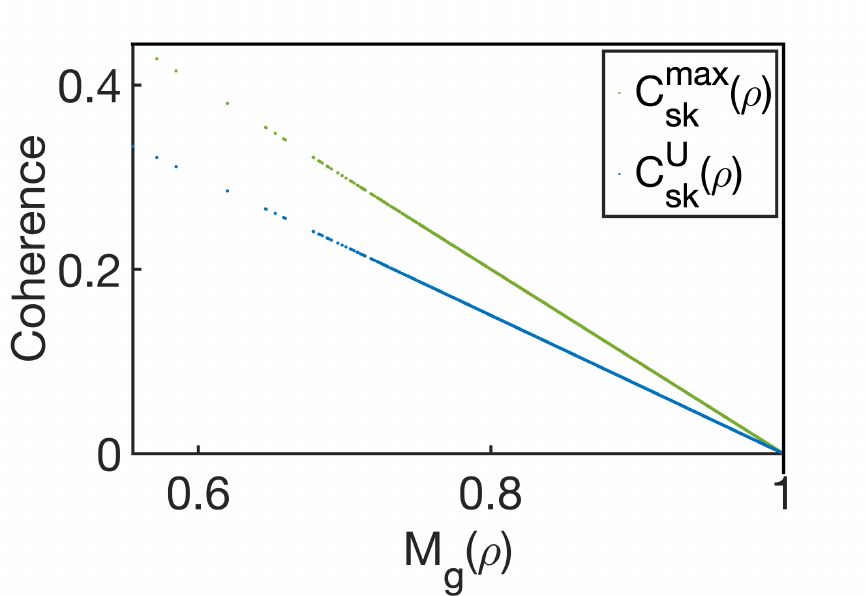}}
		\subfigure[$d=4$]{
		\centering
		\label{figex1b}
		\includegraphics[width=4.1cm]{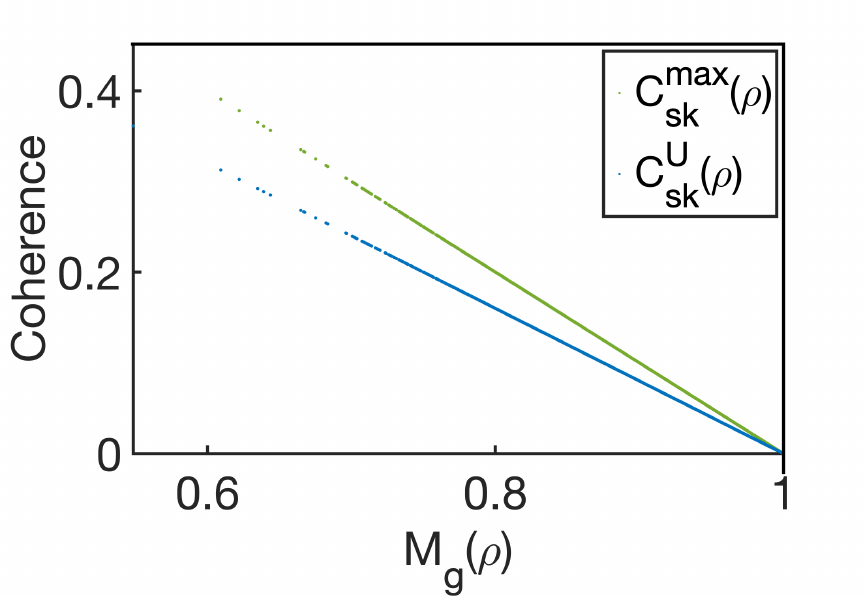}}
		\subfigure[$d=5$]{
		\centering
		\label{figex1b}
		\includegraphics[width=4.1cm]{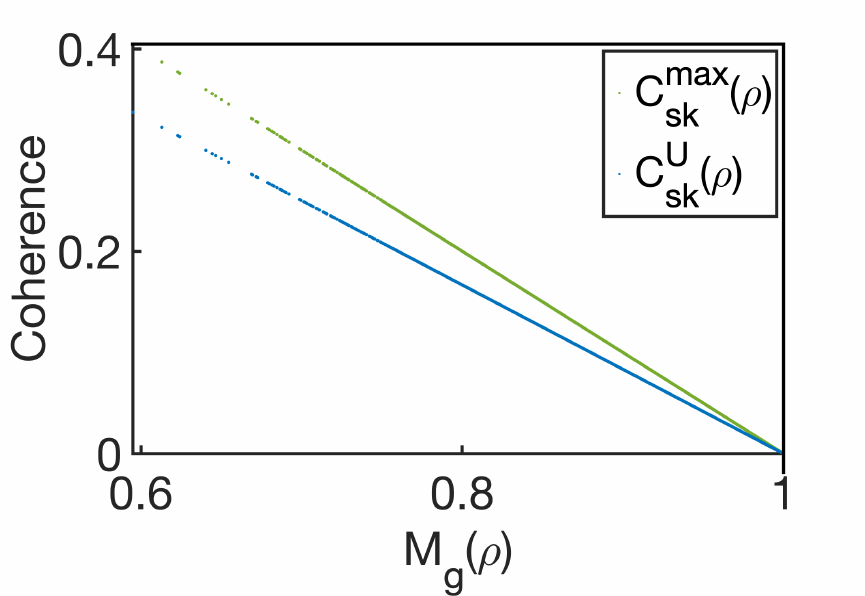}}
 \caption{$10^5$ random states are generated to illustrate the trade-offs between the Wigner-Yanase skew information-based maximal (average) coherence and the geometric mixedness.}
 \label{fig3}
 \end{figure}

According to the spectral decomposition theorem, an arbitrary qudit state can be decomposed into $\rho=\sum_{k=1}^{d}p_k|\psi_k\ra\la\psi_k|$, where $p_k$ and $|\psi_k\ra$ denote the $k$th eigenvalue and eigenvector, respectively. As shown in Ref. \cite{PhysRevA.108.012211}, any qudit state can be attained by generating a probability distribution $\{p_k\}$ and unitary operations constructed from the eigenvectors $|\psi_k\ra$. The uniform random number generator $f(0,1)$ outputs independent random numbers in $[0,1]$. On the one hand, one can effectively generate $d$ random probabilities $p_k$,
\begin{equation}
p_k=\frac{q_k}{\sum_{i=1}^d q_i},\ k=1,\cdots,d
\end{equation}
where $q_1=f(0,1)$ and $q_{k+1}=f(0,1)q_k$. In this way one obtains a set of probabilities in descent order.
On the other hand, one can randomly generate a $d\times d$ real matrix $R$ by using the random function $f(-1,1)$ within a closed interval $[-1,1]$. Based on the real matrix $R$, one constructs a random Hermitian matrix, $\tilde{R}=D+(U^\mathbb{T}+U)+i(L^\mathbb{T}+L)$, where $D$, $U$ and $L$, respectively, denote the diagonal, strictly upper and lower triangular parts of the real matrix $\tilde{R}$, $U^\mathbb{T}$ is the transpose of $U$. One can get $d$ normalized eigenvectors of the random matrix $\tilde{R}$. Consequently, the spectral decomposition of a random qudit state can be perfectly constructed. To illustrate our complementary relations, we adopt $10^5$ random states for dimensions $d=2,3,4,5$ to show the trade-off relations, see Figs. \ref{fig1}, \ref{fig2} and \ref{fig3}.

\section{Conclusion}\label{sec4}
We have utilized common information-theoretic tools including $l_p$ norm, Wigner-Yanase skew information and quantum entropies to investigate the intrinsic trade-offs between coherence and mixedness measures in a basis-independent manner. We have demonstrated rigorous constraints whereby the maximal and average coherences are fundamentally limited by the degree of mixedness under alternate quantifications.

Our trade-off relations generalize the prior basis-dependent relations \cite{CHE2023106794,PhysRevA.64.042113,PhysRevA.91.052115,Fu_2022}, which provide fundamental insights into the latent coherence resources present within arbitrary quantum systems that undergo decoherence. These relations quantify the inherent limits on extractable coherence imposed by environmental noise related to such as quantum controls. They elucidate the inherent restrictions on simultaneous coherence and mixedness within quantum systems subjected to open dynamics. The results provide impetus to the study of important physical quantities in open quantum systems and the effect of noise on quantum resources.

\bigskip
\noindent{\bf Acknowledgments}\, \,
This work is supported by the National Natural Science Foundation of China (NSFC) under Grants 12075159 and 12171044, the specific research fund of the Innovation Platform for Academicians
of Hainan Province under Grant No. YSPTZX202215, and Changsha University of Science and Technology (Grant No. 097000303923).

\bibliographystyle{apsrev4-2}
\bibliography{zhangcohmix}

\end{document}